\documentclass[aps,prl,twocolumn, preprintnumbers]{revtex4}
\usepackage{mathrsfs}
\usepackage{amsmath, amssymb, amsfonts, bbm}
\usepackage{graphics,epsfig,color,subfigure,graphicx,epsf}

\newcommand{\be}{\begin{equation}}
\newcommand{\ee}{\end{equation}}
\def\bea{\begin{align}}
\def\ena{\end{align}}

\def\nnb{\nonumber}

\def\bea{\begin{eqnarray}}
\def\eea{\end{eqnarray}}
\def\nnb{\nonumber}

\begin{document}

\preprint{UMD-40762-449}
\preprint{PUPT-2298}

\title{A bound on the speed of sound from holography}

\author{Aleksey Cherman}
\email{alekseyc@physics.umd.edu}
\affiliation{Center for Fundamental Physics, Department of Physics, University of Maryland,
College Park, MD 20742-4111}

\author{Thomas D. Cohen}
\email{cohen@physics.umd.edu}
\affiliation{Center for Fundamental Physics, Department of Physics, University of Maryland,
College Park, MD 20742-4111}

\author{Abhinav Nellore}
\email{anellore@princeton.edu}
\affiliation{Joseph Henry Laboratories, Princeton University, Princeton, NJ 08544}

\begin{abstract}
We show that the squared speed of sound $v_{s}^{2}$ is bounded from above at high temperatures by the conformal value of $1/3$ in a class of strongly coupled four-dimensional field theories, given some mild technical assumptions.  This class consists of field theories that have gravity duals sourced by a single scalar field.  There are no known examples to date of field theories with gravity duals for which $v_{s}^{2}$ exceeds $1/3$ in energetically favored configurations.  We conjecture that $v_{s}^{2}=1/3$ represents an upper bound for a broad class of four-dimensional theories.
\end{abstract}

\maketitle
{\it Introduction.}  The gauge/gravity duality, which relates gauge theories to string theories in higher-dimensional spaces~\cite{GaugeGravity}, has been used in recent years to shed light on the properties of plasmas described by strongly coupled, large $N_c$ gauge theories.  When a field theory that has a string dual is in the large $N_c$ and strong coupling limits, the string dual generally reduces to a classical supergravity theory.  This allows one to get information on strongly coupled quantum gauge theories by doing classical calculations in higher-dimensional `holographic' gravitational theories.  For instance, transport coefficients, which are normally theoretically inaccessible in strongly coupled systems, can be calculated in theories with gravity duals.

Unfortunately, there are no known gravity duals to the gauge theories that are currently used to describe nature.  In particular, there is no known gravity dual to QCD, which for $N_c=3$ describes the strongly coupled quark-gluon plasma under exploration at RHIC.  The gauge/gravity duality therefore makes no quantitative predictions for phenomenologically interesting theories.  However, one might hope to learn some qualitative lessons by searching for quantities that do not depend sensitively on the details of any particular gravity dual.  With luck, insights gleaned from such `universal' properties may extend to theories with no known duals.  A striking example of a universal quantity is the ratio $\eta/s$ of shear viscosity to entropy density. This takes the value $1/4\pi$ in all theories with gravity duals (i.e., certain field theories that are in the large $N_c$ and strong 't Hooft coupling limits)~\cite{EtaBound}.  The phenomenological implications of the universality of $\eta/s$ in theories with gravity duals remain unclear~\cite{ManySpecies}.  Moreover, the deviations from $\eta/s$ as one moves away from the supergravity limit do not appear to be universal, as they can either increase or decrease $\eta/s$~\cite{EtaBoundViolations}.  Nevertheless, the search for universal behaviors in holographic theories remains important both for theoretical reasons and for its possible phenomenological relevance.

In this paper, we show that in the simplest class of nonconformal four-dimensional ($4D$) theories with gravity duals, the squared speed of sound $v_s^2$ is always bounded from above by $1/3$ at sufficiently high temperatures.  (We work in natural units with $c = \hbar = k_{B} = 1$ throughout, and we focus on $4D$ field theories for definiteness.)   Of course,  $v_s$ is not universal in the same sense as $\eta/s$; in general it depends on the temperature $T$, the chemical potential $\mu$, and other details of a system. However, as we will show, this dependence takes a certain universal form at high $T$ in the class of theories that we consider.  Furthermore, since $v_s^2 > 1/3$ has to date never been observed in energetically stable configurations of theories with gravity duals, it is tempting to speculate that $v_s^2=1/3$ is an upper bound in a broad class of strongly coupled gauge theories.

{\it Single-scalar systems.}  The squared speed of sound can be written as $v_{s}^{2} = \partial p /\partial \epsilon$, where $p$ is the pressure of a system and $\epsilon$ is its energy density.  For systems at zero chemical potential, $v_s^2$ can also be written in terms of the entropy density $s$ as
\be
\label{vs2Formula}
v_s^2 = \frac{d \log T}{d \log s} \;.
\ee
In a $4D$ conformal field theory (at finite $T$), the entropy density $s$ behaves as $s\sim T^3$, and $v_s^2 = 1/3$. However, in nonconformal field theories, $v_s^2$ has a nontrivial dependence on the temperature and other properties of the theories.  The simplest class of gravity duals that describe nonconformal strongly coupled $4D$ field theories at large $N_c$ and zero chemical potential is the so-called `single-scalar model'~\cite{GubserNellore,GursoyEtAl,DeWolfeRosen} with the action
\be
\label{S5D}
S = \frac{1}{2 \kappa_{5}^2} \int{d^5\,x \sqrt{-g} \left[ R - \frac{1}{2}(\partial \phi)^2 - V(\phi) \right]} \;.
\ee
Above,  $\kappa_5^2/(8\pi)$ is the $5D$ gravitational constant, $\phi$ is a real scalar field, and $V(\phi)$ is a smooth potential that is symmetric about an extremum at $\phi=0$.  Different potentials correspond to different dual gauge theories.

We study systems at finite temperature with translational invariance in the $(t, \vec{x})$ directions and $SO(3)$ invariance in the $\vec{x}$ directions.  The most general metric ansatz consistent with these symmetries is
\be
\label{metricansatz}
ds^2 = a^2 (-h dt^2 + d\vec{x}^2)+ \frac{dr^2}{b^2 h}  \;,
\ee
where $a, b,$ and $h$ are functions of the holographic coordinate $r$ only, and $\phi = \phi(r)$.  A black hole horizon occurs at $r=r_h$, where $h$ has a simple zero.  $a, b$ and $\phi$ are all regular at $r=r_h$.
For this system, $s$ and $T$ can be read off in the usual way:
\be
\label{SandT}
s = \frac{2\pi}{\kappa_{5}^2} |a(r_h)|^3  \, ,\; \; \; \; \;
T = \frac{|a(r_h)b(r_h)h'(r_h)|}{4\pi} \; .
\ee

We consider theories that become approximately conformal deep in the UV.  This corresponds to assuming that the gravity dual approaches $5D$ anti-de Sitter space ($AdS_5$) as $|a| \rightarrow \infty$.  To obtain asymptotically $AdS_5$ geometries in the UV as $\phi\rightarrow 0$, $V(\phi)$ must behave as
\be
\label{asympV}
\lim_{r\rightarrow \infty} V(\phi) = -\frac{12}{L^2} + \frac{1}{2L^2} \Delta(\Delta-4) \phi^2 + \mathcal{O}(\phi^{4}) \;,
\ee
with $L$ the $AdS$ curvature radius, $\Delta(\Delta-4)/L^2$ the squared mass $m^2$ of the scalar, and $\Delta$ the scaling dimension of the gauge theory operator $\mathcal{O}_{\phi}$ dual to the scalar field $\phi$~\cite{GaugeGravity}.   We restrict our attention to relevant $\mathcal{O}_{\phi}$, for which $\Delta < 4$.  Since stable backgrounds must satisfy the Breitenlohner-Freedman bound $m^2 L^2 > -4$~\cite{BFbound}, we focus on $2 < \Delta < 4$ in our analysis.

A potential with just the constant term $-12/L^2$ would have an $AdS_5$-Schwarzschild black hole as a solution to the equations of motion with a conformal field theory (CFT) at finite $T$ living on the $AdS_5$ boundary.  Thus, a $4D$ gauge theory that has a description in terms of a $5D$ single-scalar gravity dual is just a CFT deformed by the addition of a relevant operator.   As illustrated in Ref.~\cite{GubserNellore,GursoyEtAl}, the thermodynamics of these single-scalar systems is quite rich: Many equations of state $v_{s}(T)$ are possible, and the system can undergo first- or second-order phase transitions, depending on the form of $V(\phi)$.

We now demonstrate that in all models of this class, $v_{s}^2 \le 1/3$ at high $T$.  As our derivation will self-consistently show, small $\phi_H \equiv \phi(r_h)$ corresponds to asymptotically high $T$~\cite{highTnote}, and at high $T$ these models are sensitive only to the `universal' part of $V(\phi)$ given in Eq.~(\ref{asympV}).  By high $T$, we mean temperatures that are large compared to all other scales in the problem.  Since these systems are approximately conformal in the UV, we expect that at very high $T$, the background geometry will approximate an AdS-Schwarzschild black hole.  Thus, $v_{s}^{2}$ should approach the conformal value of $1/3$ as $T\rightarrow \infty$.  However, the sign of the first nonzero correction in a high-temperature expansion of $v_{s}^{2}$ is less obvious.

To determine the sign of this correction, we solve the equations of motion resulting from Eq.~(\ref{S5D}) perturbatively around the AdS-Schwarzschild black hole solution.  Working to second order in $\phi_H$, it is possible to obtain a closed-form expression for the backreaction of the scalar field on the geometry, which then allows us to evaluate $s$ and $T$ and find $v_{s}^{2}$ using Eq.~(\ref{vs2Formula}).  We sketch the derivation in an appendix and simply present the result here:
\bea
\label{vs2_result}
 v_{s}^{2} (\phi_H) &=& 1/3 - C(\Delta) \phi_{H}^{2}+\mathcal{O}(\phi_{H}^{3}),\; \mathrm{where}\nnb \\
 C(\Delta) &=& \frac{1}{576} (\Delta-4)^2 \Delta \Big[16+(\Delta-4)\Delta  \nnb\\ 
 &\times& \, \int_{1}^{\infty} {ds\,s\, {}_{2}F_{1}(2-\Delta/4; 1+\Delta/4; 2;1 - s)^2} \,\Big] \nnb \\
&=& \frac{1}{18 \pi} (4-\Delta)(4-2\Delta) \tan \left(\pi \Delta/4\right)\;,
\eea
and ${}_{2}F_{1}$ is a hypergeometric function.  The simplified form in the last line can be obtained by standard identities. Since $C(\Delta)$ is positive, $v_{s}^{2} \le 1/3$.  This result was also obtained by different methods in Ref.~\cite{HohlerStephanov}.

\begin{figure}[t]
\centering
\includegraphics[scale=.7]{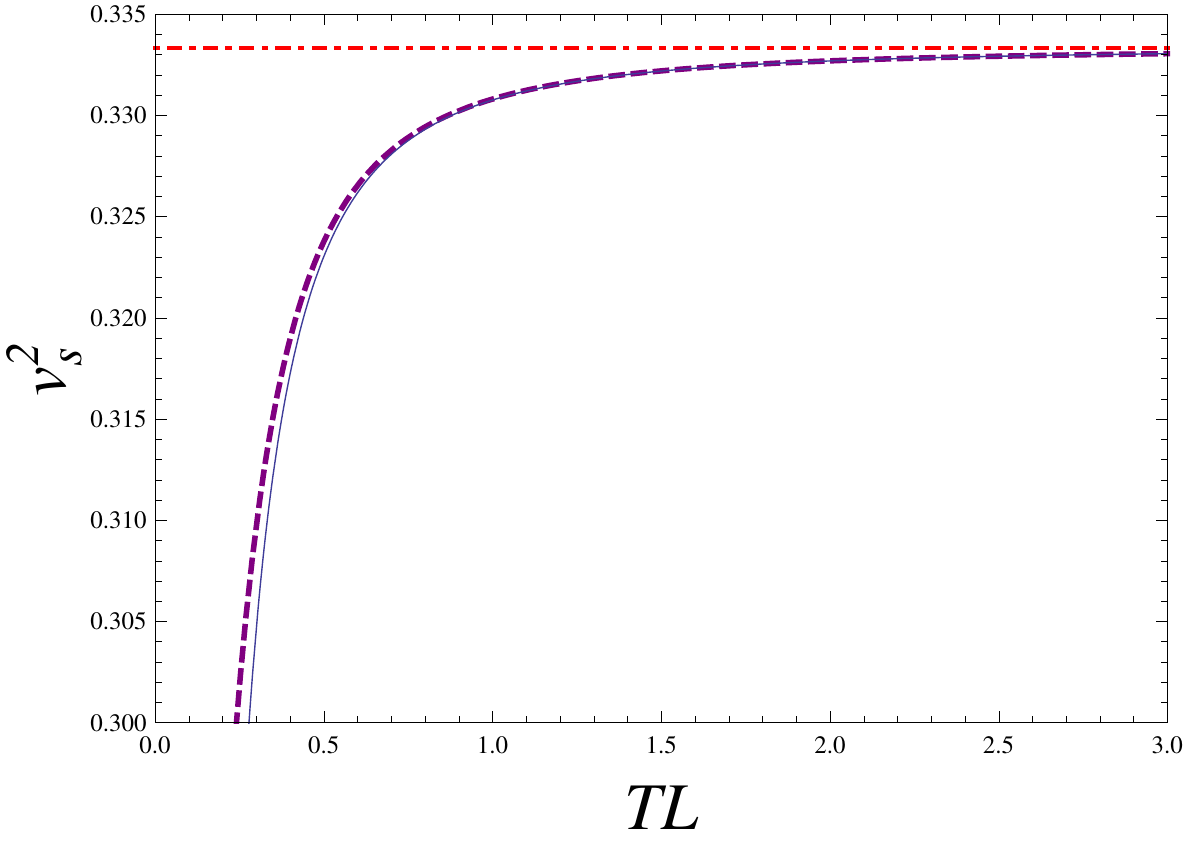}
\caption{Plot of the high-temperature approximation to $v_s^2$ in Eqs.~(\ref{vs2_result}) and  (\ref{TfromPhiH}) for $\Delta=3$ (solid line) versus a numerical solution (dashed line) for $v_s^2$ found using the methods of Ref.~\cite{GubserNellore}.  The numerical solution is for $V(\phi) = -\frac{12}{L^2} \cosh(\frac{1}{2}\phi)$, which corresponds to $\Delta=3$ . The sound bound $v_s^{2}=1/3$ is shown as a horizontal dot-dashed line.}
\label{Fig}
\end{figure}

Using the methods discussed in the appendix, it is not hard to show that
\be
\label{TfromPhiH}
\phi_H = (\pi L T)^{\Delta-4} \frac{\Gamma(\Delta/4)^2}{\Gamma(\Delta/2-1)}
\ee
plus corrections that go to zero as $T \rightarrow \infty$.  As promised, small $\phi_H$ corresponds to high temperatures.   In Fig.~\ref{Fig}, we demonstrate that the closed-form result for $v_{s}^{2}$ in Eqs.~(\ref{vs2_result}) and (\ref{TfromPhiH}) matches a numerical solution for $v_{s}^{2}$ at large $T$.  

%%%%%%%%%%%%%%
%%%%%%%%%%%%%%%

{\it Speed of sound in other models.}  The result that we have obtained above for single-scalar holographic models fits into a pattern.  All other calculations of $v_{s}^{2}$ in theories with gravity duals in the literature to date have found $v_s^2\le1/3$ at high temperatures, at least in energetically stable systems.  We review some representative results from the literature below.

The speed of sound has been calculated to be $1/5$~\cite{SSsound} in the finite-temperature generalization of the Sakai-Sugimoto model~\cite{SS}, which is dual to a strongly coupled field theory that contains fundamental matter~\cite{D3D7}.  It has also been shown that $v_s^2 \le 1/3$ at high $T$ in the D3/D7 system~\cite{D3D7sound}.

The speed of sound at high temperatures has also been computed in a $4D$ cascading gauge theory~\cite{CascadingSound}.  This strongly coupled theory has the effective gauge group $SU(K) \times SU(K+P)$ at high $T$, where $P \ll K$, and $v_s^2$ was calculated using a gravity dual to be
\be
v_s^{2}=\frac{1}{3} - \frac{4}{9}  \frac{P^2}{K}+\mathcal{O}\left(\frac{P^4}{K^2}\right)
\ee

At high $T$, $v_s^2$ has also been calculated in $\mathcal{N}=2^{*}$ gauge theory~\cite{N2Sound}.  This theory is $4D$ $\mathcal{N}=4$ super Yang-Mills theory at finite $T$, which is deformed by turning on small masses for the bosons and fermions in two of the $\mathcal{N}=1$ chiral multiplets that are part of the $\mathcal{N}=4$ gauge theory.  To leading order in the deformations, it was found that
\be
v_s^2=\frac{1}{3} -\frac{2 [\Gamma(3/4)]^4 m_f^2}{9 \pi^4 T^2}+\mathcal{O}(m_b,m_f/T^{4}) \;.
\ee

{\it Sound bound conjecture.}  From the examples above, it appears that $v_{s}^{2} \le 1/3$ at high temperatures in a much broader class of systems than just the single-scalar models, for which we were able to show a universal bound on the speed of sound at high temperatures.  It is not clear just how broad the class of theories that satisfy the bound is.  Motivated by the examples above, we conjecture that $v_{s}^{2} \le 1/3$ at high temperatures in all $4D$ theories with gravity duals, at least for energetically stable systems~\cite{ExoticVs2} at zero chemical potential.

It is not implausible that the bound holds more broadly.  In all of the examples in the literature so far where $v_s$ has been calculated in stable systems away from the high-$T$ limit, $v_s^2$ remains less than $1/3$~\cite{GubserNellore,BuchelCascadingHigherOrder}.  Thus, the sound bound $v_s^2 \le 1/3$ may continue to hold away from the high-$T$ limit for some broad class of theories.

It is a challenge for future work to determine the class of systems for which such a sound bound may apply.  Note that the sound bound cannot apply to \emph{all} field theories in nature.  For instance, in QCD at zero temperature and nonzero isospin chemical potential $\mu_{I}$, the speed of sound can be accurately calculated in chiral perturbation theory provided one works in the regime $\Lambda \gg  \mu_{I},  m_{\pi}$ (where $\Lambda$ is a typical hadronic scale and $m_{\pi}$ is the pion mass)~\cite{SonStephanov}.   In the nontrivial phase with  $ \mu_{I} >m_{\pi}$,
\be
v_s^2 = \frac{\mu_{I}^2 - m_{\pi}^2}{\mu_{I}^2 +3 m_{\pi}^2} \;,
\ee
so $v_s^2 \rightarrow 1$ as $m_{\pi}^2 \rightarrow 0$.  Similar behavior was seen earlier in more ad hoc models~\cite{WaleckaModel,Zeldovich}.

The counterexample above shows that the sound bound cannot generally apply to all systems with chemical potentials.   However, as was shown in Ref.~\cite{FiniteMuHolog}, $v_s^2 \le 1/3$ in a $D3/D7$ holographic model at finite isospin chemical potential, even when $\mu \gg T$.  Thus, it is conceivable that the sound bound may extend even to theories with gravity duals at finite chemical potentials.

There are some heuristic field-theoretic arguments that suggest that $v_s^2 \le 1/3$ at high temperatures for systems at zero chemical potential.  For example~\cite{Springer}, suppose we write the entropy in terms of the number of effective degrees of freedom $N_{\mathrm{eff}}(T)$, so that $s = \frac{16 \pi^2}{45} N_{\mathrm{eff}}(T) T^3$.  Then we have $v_s^{-2} = 3+T N_{\mathrm{eff}}'(T)/N_{\mathrm{eff}}(T)$.  In asymptotically free theories, one expects that at high temperatures $N_{\mathrm{eff}}'(T)\ge 0$, with $N_{\mathrm{eff}}'(T)\rightarrow 0$  at large $T$.   This makes it plausible that in such theories, which remain well-defined in the UV, the speed of sound will approach $1/3$ from below.   In fact, this is what happens in QCD at zero chemical potential \cite{QCDvs2}.

Of course, the heuristic argument above only applies cleanly to theories that are weakly coupled at high temperatures, so it does not apply directly to theories with gravity duals.   It was conjectured in Ref.~\cite{AppelquistEtAl} that for strongly coupled theories, $N_{\mathrm{eff}}(T)$ in the UV must be greater than in the IR.  However, this by itself does not imply the sound bound unless $N_{\mathrm{eff}}(T)$ grows monotonically, which is not always the case.

{\it Conclusions.}  We have shown that $v_s^2 \le 1/3$ at high temperatures in theories with single-scalar gravity duals.  Our techniques can be extended to show that $v_s^2 \le1/3$ in systems with multiple scalars and also to compute transport coefficients other than $v_s^2$, as will be discussed in a companion paper \cite{ChermanNellore}.  It would be interesting to study other strongly coupled systems with gravity duals to determine what class of systems satisfies the sound bound of $v_s^2 \le 1/3$, and to see if there are classes of systems where such a bound can apply beyond just the high-temperature limit.  To investigate whether there is a class of field theories without gravity duals for which such a bound might apply, it would be useful to explore the $1/N_c$ and $1/\lambda$ corrections to $v_s^2$ in gauge/string duality.

{\it Acknowledgements.}  We thank P.~Hohler and M.~Stephanov for pointing out that the integral over hypergeometric functions in Eq.~\ref{vs2_result} can be simplified, and for bringing their related work \cite{HohlerStephanov} to our attention after this work was completed. We thank P.~Hohler for discussions at an early stage of this work.  We thank P.~Bedaque, S.~Gubser, M.~Kaminski, P.~Kerner, F.~Rocha, T.~Springer, E.~Shuryak,  D.~T.~Son, M.~Stephanov, and B.~Tiburzi for useful discussions.  The work of A.~C. and T.~D.~C was supported by the US DoE under grant DE-FG02-05ER41368, and the work of A.~N. was supported by the US NSF under grant PHY-0652782.

{\it Appendix. }  Consider the action Eq.~(\ref{S5D}) with $V(\phi)$ as in Eq.~(\ref{asympV}) and the metric ansatz Eq.~(\ref{metricansatz}).  There are three independent equations of motion:  the scalar equation and the $tt$ and $rr$ components of Einstein's equations.  Let us work in the gauge $a=r$ so that $r\rightarrow\infty$ at the boundary.  When $\phi$ vanishes identically, the general solution to the equations of motion is $AdS_{5}$-Schwarzschild.  When $\phi$ is everywhere small, it assumes the profile
\be
\label{firstscalar}
\phi(r) = \phi_0 \,{}_{2}F_{1}(1-\Delta/4;\Delta/4;1;1-r^4/r_h^4) \;,
\ee
where $\phi_0$ is an integration constant that measures the smallness of the scalar.  Eq.~(\ref{firstscalar}) is one solution to the linearized scalar equation of motion in $AdS_{5}$-Schwarzschild.  We have discarded a second solution that is not regular at $r=r_h$, the location of the black hole horizon.

Now imagine fixing the entropy density $s$ of $AdS_{5}$-Schwarzschild at some large $s_0$ in the high-temperature regime.  We need to find the temperature that corresponds to $s_0$ when $\phi$ is turned on.  To this end, we pursue perturbation expansions of the metric and the scalar in powers of $\phi_0$.  Working up to $\mathcal{O}(\phi_0^n)$ is sufficient for computing corrections to $v_s^2$ up to $\mathcal{O}(\phi_0^n)$.  The metric backreacts on the scalar at odd orders in $\phi_0$, and the scalar backreacts on the metric at even orders in $\phi_0$.  Meanwhile, the dual CFT lagrangian has been deformed by the addition of the relevant term $\Lambda^{4-\Delta}\mathcal{O}_{\phi}$.  Here, $\Lambda$ is a new energy scale that also appears in the leading behavior of the scalar at the boundary:
\be
\label{phinearbound}
\phi\approx(\Lambda L)^{4-\Delta}r^{\Delta-4}.
\ee
We must keep $\Lambda$ fixed when computing equations of state, so we set $\Lambda L=1$.  Comparing the asymptotic form of Eq.~(\ref{firstscalar}) as $r\rightarrow\infty$ with Eq.~(\ref{phinearbound}) then gives rise to a relationship connecting $r_h$ and $\phi_0$:
\be
\label{rhrelationship}
r_{h}^{\Delta-4}\Gamma(\Delta/4)^2=\phi_0\Gamma(\Delta/2-1).
\ee
Fixing $\Lambda$ thus imposes a consistency condition on the expansion.  Small $\phi_0$ necessarily corresponds to large $r_h$, which means that the exact single-scalar background only approaches $AdS_5$-Schwarzschild in the `conformal' high-temperature limit.  The four boundary conditions we impose at each order in $\phi_0$: (1) maintain $\Lambda L=1$ and hence also Eq.~(\ref{rhrelationship}); (2) retain the horizon location $r=r_h$ so that $s$ remains at $s_0$; (3) ensure that the solution is regular at $r=r_h$; and (4) preserve the boundary asymptotics $h\rightarrow 1$ and $b\rightarrow r/L$.  We do not present our results for $b$ and $h$ here because they are cumbersome, but we note that $\phi_H=\phi_0$ up to $\mathcal{O}(\phi_0^2)$.  $v_s^2$ is computed straightforwardly to $\mathcal{O}(\phi_H^2)$ by plugging the $\mathcal{O}(\phi_0^2)$ results for $b$ and $h$ into Eqs.~(\ref{SandT}), eliminating $r_h$ using Eq.~(\ref{rhrelationship}), and subsequently applying the formula Eq.~(\ref{vs2Formula}).


\begin{thebibliography}{99}
\bibitem{ExoticVs2}
In Ref.~\cite{BuchelPagnuttiExotic} $v_s^2$ was seen to exceed the conformal value in energetically unstable thermodynamic phases that somewhat exotically persist to arbitrarily high temperatures.  However, the energetically stable phase in that system has $v_s^2$ bounded from above by the conformal value.
\end{thebibliography}
\end{document}